%% file: main.tex
\documentclass[conference]{IEEEtran}

\usepackage{paralist}
\usepackage[]{todonotes}
\usepackage[noadjust]{cite}
\usepackage{xspace}
\usepackage{amsmath}
\usepackage{multirow}
\usepackage{balance}
\usepackage{makecell}
\AtBeginDocument{
  \definecolor{pdfurlcolor}{rgb}{0,0,0.6}
  \definecolor{pdfcitecolor}{rgb}{0,0.6,0}
  \definecolor{pdflinkcolor}{rgb}{0.6,0,0}
  \definecolor{light}{gray}{.85}
  \definecolor{vlight}{gray}{.95}
}
\usepackage{url} 
\usepackage[colorlinks=true,citecolor=pdfcitecolor,urlcolor=pdfurlcolor,linkcolor=pdflinkcolor,pdfborder={0 0 0}]{hyperref}
\newcommand{\simgrid}{SimGrid\xspace}

\renewcommand{\arraystretch}{1.03}
\begin{document}

\title{Automated Calibration of Parallel and\\Distributed Computing Simulators: A Case Study}

\makeatletter
\newcommand{\linebreakand}{%
  \end{@IEEEauthorhalign}
  \hfill\mbox{}\par
  \mbox{}\hfill\begin{@IEEEauthorhalign}
}
\makeatother

\author{\IEEEauthorblockN{Jesse McDonald}
	\IEEEauthorblockA{
		\textit{University of Hawai`i at M\=anoa}\\
		Honolulu, HI, USA \\
		jamcd@hawaii.edu}
	\and
	\IEEEauthorblockN{Maximilian Horzela}
	\IEEEauthorblockA{
		\textit{Karlsruhe Institute of Technology}\\
		Karlsruhe, Germany \\
		maximilian.horzela@kit.edu}
	\linebreakand
	\IEEEauthorblockN{Frédéric Suter}
	\IEEEauthorblockA{
		\textit{Oak Ridge National Laboratory}\\
		Oak Ridge, TN, USA \\
		suterf@ornl.gov}
	\and
	\IEEEauthorblockN{Henri Casanova}
	\IEEEauthorblockA{
		\textit{University of Hawai`i at M\=anoa}\\
		Honolulu, HI, USA \\
		henric@hawaii.edu}
}

\maketitle

\begin{abstract}
\input{abstract}
\end{abstract}
\input{sec-intro}
\input{sec-related}
\input{sec-approach}

\input{sec-casestudy}

\input{sec-conclusion}


\section*{Acknowledgments}
\noindent
This work was partially supported by NSF Awards \#2106059 and \#2103489, the German Federal Ministry of Education and Research (project FIDIUM 05H21VKRC2) and the Institute of Experimental Particle Physics (ETP) at the Karlsruhe Institute of Technology, Germany. The technical support and advanced computing resources from University of Hawai`i Information Technology Services - Cyberinstructure, funded in part by the NSF MRI Award \#1920304, are gratefully acknowledged.


\balance
\bibliographystyle{IEEEtran}
\bibliography{biblio}

\end{document}

%% file: abstract.tex
Many parallel and distributed computing research results are obtained in simulation, using simulators that mimic real-world executions on some target system.  Each such simulator is configured by picking values for parameters that define the behavior of the underlying simulation models it implements.  The main concern for a simulator is accuracy: simulated behaviors should be as close as possible to those observed in the real-world target system. This requires that values for each of  the simulator's parameters be carefully picked, or ``calibrated," based on ground-truth real-world executions. Examining the current state of the art shows that simulator calibration, at least in the field of parallel and distributed computing, is often undocumented (and thus perhaps often not performed) and, when documented, is described as a labor-intensive, manual process.  In this work we evaluate the benefit of automating simulation calibration using simple algorithms.  Specifically, we use a real-world case study from the field of High Energy Physics and compare automated calibration to calibration performed by a domain scientist.  Our main finding is that automated calibration is on par with or significantly outperforms the calibration performed by the domain scientist.  Furthermore, automated calibration makes it straightforward to operate desirable trade-offs between simulation accuracy and simulation speed.

%% file: sec-intro.tex
\section{Introduction}
\label{sec:intro}

Much Parallel and Distributed Computing (PDC) research relies on
experimental results obtained by executing application workloads on PDC
platforms.  Many published works include results obtained in simulation,
either in addition to or as a replacement for results obtained from
real-world experiments.  Simulation is attractive for several reasons. It
makes it possible to explore hypothetical workload and platform
configurations.  It also yields results that are 100\% reproducible
and observable.  Finally, in most cases, simulation experiments entail
significantly less time, labor, carbon footprint, and/or funds than
their real-world counterparts.  The main concern with simulation
is \emph{accuracy}, that is, how well real-world and corresponding
simulated executions match.

Consider a PDC system that consists of some software stack used to execute
an application workload on a hardware platform,  and a simulator of this
system. The simulator's underlying \emph{simulation models} can operate at
various levels of abstraction.  As an example consider a distributed file
system whose performance needs to be studied in simulation.  A simulator
could be developed that simulates the system as a black box server with
some stochastic service time (high level of abstraction), or as a set of
daemons that perform I/O operations on local disks and coordinate over the
network using some distributed algorithm (low level of abstraction).
Whenever data is sent to the distributed file system the time necessary for
that data transfer could be based on packet-level simulation of network
communications (low level of abstraction) or computed as a data size
divided by a bandwidth (high level of abstraction).  While low levels of
abstractions can potentially lead to high simulation accuracy they also
typically increase the space- and time-complexity of the simulator, hence
the well-recognized trade-off between simulation accuracy and simulation
scalability (speed and memory footprint).

Regardless of their levels of abstraction, the simulation models in a PDC
simulator all come with configuration \emph{parameters}.  These parameters
pertain to the hardware platform (e.g., network and disk bandwidths, CPU
clock rates, cache sizes), the software stacks (e.g., TCP window size,
software overhead to start a virtual machine instance, size of some control
messages), and the application workload (e.g., data and compute volumes, task
granularity, control/data dependencies). Simulation models can be developed
from scratch by simulator developers themselves, who then define the set of
relevant parameters.  Alternately, simulator developers can use simulation
models provided by simulation frameworks designed specifically to ease PDC
simulator development~\cite{gridsim, CloudSim, GroudSim, DISSECTCF, CODES,
SST, iFogSim, casanova2020fgcs, casanova2014simgrid}.

Different values for the parameters of the simulation models used by a
simulator of a target PDC system lead to different simulated executions.  It
is thus critical that values be chosen that make the simulation is as
accurate as possible.  Some parameter values may be straightforward to
determine, such as the parameters that define the application workload.
But for others, typically those that pertain to the hardware
platform, picking good values can be challenging.  A seemingly natural
approach is to pick values based on knowledge of the PDC system.
Unfortunately, these values may not be known to the user (e.g., parameters
that define the precise network topology).  Furthermore, if the simulator's
level of abstraction is high, a single parameter value may not map directly
to a single known characteristic of the target system.  For instance, if a
simulator abstracts a complicated network topology as a single network link
with a latency and a bandwidth, it is difficult to come up with reasonable
values for these two parameters based on knowledge (assuming this knowledge
even exists) of the real-world network topology.  A more sound approach is
instead to pick parameter values that maximize simulation accuracy with
respect to one or more execution scenarios on a real-world system.  We term
this approach \emph{simulation calibration}. We say that simulation models,
or simulators that use these models,  have been calibrated if all parameter
values have been determined so that simulated executions are as similar as
possible to \emph{ground-truth} real-world executions. The hope is that the
calibrated simulator will then achieve high accuracy when simulating
executions that go beyond the ground-truth executions used to calibrate it
(different application workloads, different platform scales, different
platform hardware characteristics, etc.).

Simulation calibration ensures that meaningful conclusions can be drawn
from simulation results obtained with PDC simulators.  Yet, many research
works present simulation results without any mention of calibration,
perhaps indicating that it was not performed. This may be because these
works use simulators developed using some PDC simulation framework, which
come with built-in simulation models. These models typically come with
default parameter values, and these default values may have been obtained based
on calibration for some particular real-world system (or averaged over
calibrations for multiple real-world systems). At any rate, there is no
guarantee that these default values will be appropriate for all possible
scenarios. Some works do mention calibration and give some details
about the calibration procedure, which is typically a labor-intensive, at
least partially manual, procedure.

We claim that, overall, calibrating simulators of PDC systems is challenging
and often not performed (sufficiently or sufficiently thoroughly) in practice.
In this work, we verify this claim based on inspection of the literature, 
and evaluate the potential benefit of automating simulation calibration.
More specifically, our contributions include: 

\begin{compactitem}
    \item An investigation of the state of the art of the calibration of
          simulators of PDC systems;
    \item An instantiation of a general automated simulation calibration procedure with
          simple algorithms;
    \item An evaluation of the benefit of automated simulation
          calibration when compared to a calibration performed by a domain scientist
	  for a production use case. 
\end{compactitem}

This paper is organized as follows. Section~\ref{sec:related} reviews
related work and attempts to characterize the state of the art of
PDC simulator calibration. Section~\ref{sec:approach}
defines the simulation calibration problem and describes our
algorithms. Results are presented in
Section~\ref{sec:casestudy} for a High Energy Physics production use case.  Finally,
Section~\ref{sec:conclusion} concludes with a brief summary and
perspectives on future work.

%% file: sec-related.tex
\section{Related Work}
\label{sec:related}

\subsection{Simulation of PDC Systems}
\label{sec:related:simulation}


Many frameworks have been developed for PDC system simulation. 
Several have garnered sizable user communities and are still actively
being maintained at the time of writing~\cite{gridsim, CloudSim, GroudSim,
DISSECTCF, CODES, SST, iFogSim, casanova2020fgcs, casanova2014simgrid}.
Different frameworks achieve different compromises between accuracy and
scalability by implementing different kinds of simulation models.  At one
extreme are models that are designed at low levels of abstraction to
capture ``microscopic'' behaviors of hardware/software components, which
favors accuracy over scalability (e.g., packet-level network simulation,
cycle-accurate CPU simulation, block-level disk simulation).  At the other
extreme are analytical models that abstract away microscopic behavior and
instead attempt to capture ``macroscopic'' behaviors via empirical
mathematical models.  While the latter models have lower space- and
time-complexity, they must be developed with care so that high accuracy can
be achieved~\cite{velhoTOMACS2013}.

This work is agnostic to the simulation framework used to develop the
simulator. However, if the simulation models it provides have inherent
inaccuracies (e.g., too high a level of abstraction, implementation/design
flaws), calibrating parameters optimally could still lead to low simulation
accuracy.  In Section~\ref{sec:casestudy}, we employ a simulator developed
using WRENCH~\cite{casanova2020fgcs,wrench_web}, which implements
high-level simulation abstractions for easy development of simulators of
PDC systems. WRENCH itself builds on top of
\simgrid~\cite{casanova2014simgrid,simgrid_web}, which comes with
implementations of simulation models that are high-level enough to achieve
high simulation scalability. These simulation models have also been
thoroughly validated~\cite{velhoTOMACS2013, simutool_09, nstools_07,
simgrid_storage, SMPI_TPDS,  7885814, 8048921,
7384330,stanisic,Cornebize-cluster19, casanova2020fgcs}, meaning that they
can lead to high simulation accuracy provided their configuration parameter
values are chosen appropriately.

\subsection{Calibration of PDC Simulators}

\renewcommand{\arraystretch}{1.5}
\begin{table*}
  \centering
\caption{Examination of 114 research publications in the 2017-2022 time period that include results obtained with \simgrid. }
\label{tab:simgrid_usage}
	\begin{tabular}{|p{7.7cm}|c|p{6.2cm}|c|} 
	\cline{1-2}
	\# Publications that only include simulation results & 85 & \multicolumn{2}{l}{~}\\
	\hline
	                                                                                     &                     & No comparison thereof & 4 \\\cline{3-4}
	\# Publications that include both simulation and real-world results & 29 &  Calibration perhaps performed  or at best mentioned & 15\\
       \cline{3-4}
							       &  & Calibration performed and documented & 10\\
      \hline
\end{tabular}
	\vspace*{-0.1in}
\end{table*}

The lower a simulation model's level of abstraction, the more directly its
parameters map to the characteristics of the system to be simulated.
Low-level simulation abstractions are the norm in several fields
such as computer architecture and networking. For instance,
many published networking research results are obtained using packet-level
simulators in which the lifecycle of each individual network packet is simulated
via several, and possibly many, discrete events. The parameters of the
such simulation models map directly to the physical characteristics of the network
links and routers and to the network protocol implementations in the
target real-world system to be simulated.  It
should thus be possible to pick appropriate values for these parameters
based solely on the specification of the target system. However, many
authors have found that doing so is challenging, and that calibration is
necessary for achieving high
accuracy~\cite{andel_computer06,opnet_modeler, 10.5555/1416222.1416264,
4498213, DBLP:conf/networking/HurniB09}. 


Hundreds of PDC research works have been published that include results
obtained with simulators built using various simulation frameworks.  Most
of these frameworks do not implement low-level simulation models.
Instead, they
implement simulation models at high levels of abstraction to achieve high
scalability, i.e., to make the simulation of large-scale and/or
long-running execution scenarios feasible.  Because of these high levels of
abstraction picking appropriate values for simulation model parameters is
even more challenging. That is, there may be no one-to-one correspondence
between simulator parameters and system characteristics,  meaning that even
perfect knowledge of the target system may not be sufficient to pick
appropriate parameter values.  Instead, parameter values should be
determined base on calibration with respect to ground-truth real-world
executions.

Assessing the state of the art of calibration of PDC simulators is
difficult.  But a popular PDC simulation framework,
\simgrid~\cite{simgrid_web}, maintains a list of research publications that
include results obtained using this
framework (\url{https://simgrid.org/usages.html}).  At the time of
writing, this list includes 610 publications, 114 of which are
peer-reviewed journal or conference/workshop publications for the 2017-2022
6-year time period. As an attempt to assess the state of the art, we have
examined these 114 publications in detail to determine how they perform
simulator calibration (many authors refer to calibration as ``parameter
picking" or ``parameter tuning").  Our results are summarized in
Table~\ref{tab:simgrid_usage}. Of the 114 publications, 85 include only
simulation results, which likely indicates that calibration was not
performed. In some of these works the goal is only to simulate a simplified
model of an abstract system. In others, real-world data is used as input to
the simulation, but no comparison with the real-world execution that has
generated that data is done (or can be done).  In yet other works, the goal
is to simulate a system with hardware/software technology that does not
(yet) exist. One reason why many papers do not include real-world results
is because simulation is often used precisely because such results cannot
be obtained in practice. In several of these works, however, the simulation
is intended to be representative of real-world systems. If calibration was
not performed it is not clear what these systems are.

29 of the 114 works we have reviewed include both simulation and real-world
results.  We use a broad definition of the term ``real-world", which
includes not only results obtained on real-world hardware platforms, but
also results obtained using emulation and results obtained using low-level
simulation (e.g., results obtained  using packet-level simulation of
networks).  25 of these 29 works perform or allow comparison of simulation
and real-world results.  15 of these 25 works either do not detail any
calibration procedure or merely mention that picking better simulation
parameters improves accuracy.  Some of these works, however, present
simulation results that exhibit high accuracy, which may indicate that
calibration was performed even if not mentioned.

Overall, out of the 114 publications we reviewed only 10 are explicit about
performing some calibration and give some details. Half of these describe
manual painstaking procedures by which simulation parameter values are
picked based on quantitative and qualitative comparisons of real-world and
simulation execution logs and metrics, and sometimes on inspecting the
source code of the target system's software stacks. The other half do
perform similar procedures but also rely on simple statistical techniques
(i.e., regressions).  It is important to
note that, for 8 of these 10 works, the main research contribution is a
novel simulation model. Calibration is thus necessary to validate this
simulation model. In the end, among the 106 publications that target a
non-simulation-related research topic, we found only 2 that performs a
solid and documented calibration procedure so as to ensure that simulation
results are accurate.

The above discussion indicates that simulator calibration is likely not performed
routinely in the PDC field.  Parameters may be picked based on best guesses
or simply by using the default values for models provided by simulation
frameworks.  These defaults can come from calibration with respect to some
real-world systems available to the developers of the simulation framework
(as it is the case with \simgrid).  Consequently, published simulation
results obtained using default parameter values may be valid for some
system configurations, but not necessarily for that of the particular system
of interest.  Furthermore, not all simulation models are provided by the
simulation framework, and custom models are also developed for each
particular simulator. These custom models may not come with any
(calibrated) default parameter values. Finally, we find that those works
that perform simulation calibration typically employ labor-intensive,
partially manual, procedures.

All the above provides a strong motivation to automate PDC simulator
calibration, which, to the best of our knowledge, has not been reported in
the PDC literature. The general ideal of simulation model parameter
calibration is of course not new, and has been studied from theoretical
standpoints~\cite{doi:10.1177/154851290500200405}.  It is thus not
surprising that automated calibration approaches have been proposed in many
disciplines~\cite{liu-batelaan-smedt-2005,YANG20161220,hourdakis2003practical}.


%% file: sec-approach.tex
\newcommand{\grid}{\textsc{Grid}\xspace}
\newcommand{\random}{\textsc{Random}\xspace}
\newcommand{\fixed}{\textsc{GDFix}\xspace}
\newcommand{\dynamic}{\textsc{GDDyn}\xspace}
\newcommand{\human}{\textsc{Human}\xspace}

\section{Automated Calibration}
\label{sec:approach}

\subsection{Problem Statement}
\label{sec:pb}

We define a \emph{PDC system} as: (i)~a hardware platform with 
compute and I/O resources distributed over a network; (ii)~an application
workload that consists of compute tasks that use and produce data
items; and (iii)~a runtime system that is used to execute the workload
on the platform.  Consider a real-world such system on which the
application workload has been executed repeatedly and perhaps for 
different configurations of the system (e.g., different subsets of the
resources, different application workload instances, different runtime
system configurations). Each such execution produces a \emph{ground-truth
execution trace}, i.e., a log of time-stamped execution events, such as
compute task start times and completion times.

Consider now a simulator of the system that implements several simulation
models each of which can be configured via parameters.  In practice
parameters can take many possible values (e.g., the bandwidth of a network
link in MBps, the compute speed of a core in GHz, the maximum number of
supported concurrent connections outgoing from a data server) or only a few
(e.g., a binary value that specifies whether some feature of the runtime
system is enabled).  The simulator is implemented so that it takes as input
a set of parameter values and produces an execution trace that is
comparable to a ground-truth execution trace. This comparison is done via a
user-defined metric that quantifies the discrepancy between a simulated and
a ground-truth execution trace as a measure of simulation accuracy.  The
simplest simulation accuracy metric is the relative makespan difference,
that is, the relative difference between the time elapsed between the start
of the first task and the completion of the last task in the
ground-truth and in simulation.


Given a PDC system, a set of ground-truth execution traces, a simulator of
the system, and a simulation accuracy metric, calibration is the
optimization problem that consists in determining the simulator's parameter
values that optimize the simulation accuracy metric.

The simulator developer must specify the range of possible  values of each parameter.
The narrower these
ranges the more constrained the search space, but the higher the risk that
the best parameter value lies outside that range. Simulator developers specify ranges
based on their best guesses and knowledge of the target system, but in
practice some parameter ranges could be large.  For all results in this
work we use a logarithmic representation of each parameter.  That is, each
parameter with a user-specified range $[a, b]$ is written as $2^{x}$ and
$x$ is sampled in the interval $[\log_2 a, \log_2 b]$.  The rationale is
that by sampling values logarithmically, we ensure a bigger diversity of
orders of magnitudes within the parameter range. For instance, consider a
parameter that describes the bandwidth of some network link.  Sampling
values of, say, 10~Mbps and 11~Mbps is likely useful, but sampling values
of, say, 100,000~Mbps and 100,001~Mbps probably is not.

Evaluating the objective function entails executing the simulator with 
sets of candidate parameter values, but the simulator's execution time is
non-zero and could be relatively large.  For this reason we assume that
there is a fixed bound $T$ on the time allotted to the calibration
procedure. We use a time bound rather than a bound on the number of
simulator  invocations because the value of some parameters can impact the
simulator's space- and time-complexity.

\subsection{Calibration Algorithm Implementations}
\label{sec:algs}



Many algorithms can be used for solving the simulation calibration problem.
These include simple searches, standard optimization algorithms such as
gradient descent, genetic algorithms, or Machine Learning algorithms such
as Bayesian optimization. Our goal in this work is not to determine which
level of algorithm sophistication is sufficient.  The answer to this
question is likely highly dependent on the use case at hand (simulated
system, simulator execution time, number of parameters to calibrate).
Instead our primary goal is to determine if even simple optimization
algorithms can improve upon the state of the art of manual simulator
calibration. We consider the following such algorithms:
\begin{itemize}
\item {\bf Grid Search } (\grid) --
This algorithm evaluates all parameter combinations by subdividing the
parameter space evenly in each parameter range. As the number of
subdivisions is not known in advance, each time all current subdivisions of
the range have been sampled, a new set of points to sample is determined
using the midpoints between each pair of already sampled points.  The
initial ranges are the parameter value bounds provided by the simulator developer. Thus,
given $p$ parameters to calibrate, each parameter can take one of
approximately $\sqrt[\uproot{3}p]{N}$ (evenly spaced) values in its range,
where $N$ is the total number of simulation invocations completed before the
time bound $T$ has been reached.
\item[]
\item{\bf Random search} (\random) -- This algorithm simply evaluates
sets of random parameter values, where each value is sampled uniformly
in its parameter range.
\item[]
\item {\bf Gradient Descent --} This algorithm uses a random
starting point in the parameter space. At each iteration the gradient is
approximated by sampling points a distance $\delta$ away along each
dimension. A standard backtracking line search is then used to compute the
``learning rate," i.e.,  by how much to move along the gradient to
determine the next point that should be sampled.  When the change in the
objective function between two iterations is less than~$\epsilon$, the
current search path is terminated, and a new starting point is randomly
selected.  For all results in this paper we use $\delta=0.0001$ and
$\epsilon=0.01$.  For completeness, we did consider two variations of this
algorithm:
\begin{compactenum}
	\item {\bf Dynamic} (\dynamic) -- At each iteration the value of~$\delta$ is updated to be the learning rate determined by the backtracking line search;
    	\item {\bf Fixed} (\fixed) -- The value of $\delta$ remains constant regardless of the learning rate. 
\end{compactenum}
In all our experimental results these two variants lead to almost always identical
simulation accuracy. Hence, in all that follows the results for \dynamic are omitted.
\end{itemize}

In all the above, random numbers are generated using a pseudo-random number
generator seeded with the same seed.  In our experiments all algorithms use
the same bound $T =  6$ hours. Each algorithm executes one simulation on each core of a
dedicated 2.5GHz Intel Xeon Gold 6248 40-core CPU.

%% file: sec-casestudy.tex
\newcommand{\platform}{WLCG\xspace}
\newcommand{\hitrate}{ICD\xspace}
\newcommand{\HDDcFiber}{\textsc{SCFN}\xspace}
\newcommand{\RAMcFiber}{\textsc{FCFN}\xspace}
\newcommand{\HDDcDial}{\textsc{SCSN}\xspace}
\newcommand{\RAMcDial}{\textsc{FCSN}\xspace}

\begin{figure*}[h]
    \centering
    \includegraphics[width=0.60\linewidth]{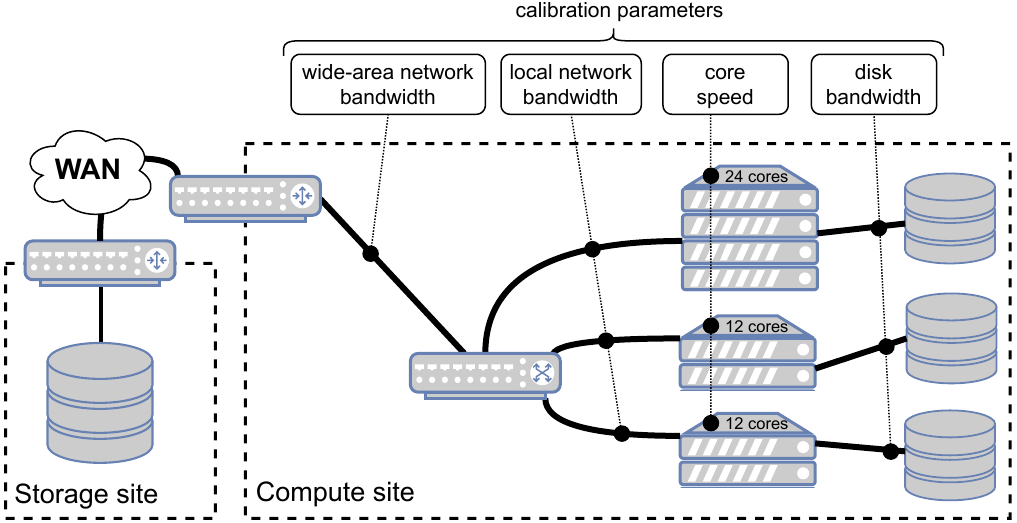}
	\caption{Execution platform.}
    \label{fig:platform}
\end{figure*}

\section{Case Study}
\label{sec:casestudy}

\subsection{Context and Objective}
\label{sec:casestudy:context}

In this section we present a case study for an application in the
field of High Energy Physics (HEP). Distributed computing platforms
are used to support the high compute and storage demands of many HEP
applications, for processing data generated by the Large Hadron Collider
(LHC) experiments and simulations.
Specifically, we consider the processing of data
generated by LHC experiments conducted for the Compact Muon Solenoid
(CMS) collaboration~\cite{CMS}, which, in 2022, required 
$\sim$415PB of tape storage and more than 1.94 billion CPU-hours. The
generated data, which describes particle collision events, can be split
into chunks that can be processed and stored independently of each other.
The processing of one event
entails multiple data reduction/transformation
steps until a final analysis step produces an output that can be stored
on a single computer and analyzed to generate scientific results. This
data processing workload is performed on a multi-site distributed
computing platform, the Worldwide LHC Computing Grid (\platform)~\cite{wlcg},
using various software infrastructures, such as HTCondor~\cite{htcondor}
for distributing computation and  XRootD~\cite{xrootd} for distributing storage.

Researchers need to estimate the execution times of current HEP
workloads of interest on (subsets of) \platform to plan experiments, to
explore various hardware resource provisioning options, and to ensure that
future CMS workloads can achieve acceptable performance.  A key performance
driver of scientific distributed computing applications is data locality,
and HEP workloads are no exception. XRootD, which is deployed on \platform,
makes it possible to deploy data caches (called ``proxy storage services")
that can perform in-memory or on-disk caching. CMS researchers need to
compare different cache deployment options in terms of the performance
boost that caching can bring to current and future workloads. The main
objective is thus to explore the large design space of combinations of
hardware resource provisioning, cache deployment, and scheduling options,
as well as workload configurations.  Achieving this objective via
real-world experiments would be too resource-consuming, especially since
\platform is used daily in production for running critical workloads. Also,
real-world experiments cannot be used to explore hypothetical (future)
scenarios. As a result, this objective can only be achieved by conducting
simulation experiments.

\subsection{Methodology}
\label{sec:methodology}

\noindent
{\bf Simulator --}
We have developed a simulator~\cite{simulator} in C++ using the WRENCH and
\simgrid simulation frameworks (see Section~\ref{sec:related:simulation}).
The simulator takes as input a description of a workload to execute and of
the \platform platform on which to execute it.  A \emph{workload} consists
of a set of independent jobs, where each job consists in reading input
files of given sizes, performing some volume of computation per byte of
input, and writing an output file of a given size.  The user can specify
data and compute volumes either as constant values or as 
probability distributions from which values are sampled.  A \emph{hardware
platform} consists of multiple sites interconnected over a wide-area
network. One or more of these sites hosts a storage service that
stores all initial input data for all jobs.  Each site comprises multi-core
compute nodes, each of which can use its local disk to cache input data.
The simulator takes as input a number between 0 and~1, called the
\emph{\hitrate} (Initially Cached Data), that denotes the fraction of input
files that are initially stored in these caches.  These compute nodes are
interconnected via a local network.

\medskip
\noindent{\bf Ground Truth Data --} Ground-truth data was obtained
with a workload that comprises 48 jobs, where each job takes 20 files as
input, each of size of $\sim$427MB. This workload was executed on
\platform using one compute site and a remote storage size, interconnected
together via a wide-area network.  The compute site hosts three 
compute nodes that are homogeneous, but two of these nodes have 12 cores while the third one has 24
cores.  All three nodes host a local HDD cache, and are connected together
via a local network. This platform configuration is depicted 
in Figure~\ref{fig:platform}. The workload was executed for \hitrate values ranging from 0
to 1 in 0.1 increments.

Each execution of the workload on the platform was conducted for 4
different configurations of the hardware platform.  Specifically, two different
network interfaces can be used for the compute site to connect to the
remote storage site (1~Gbps or 10~Gbps), and at each compute node the use
of an in-RAM disk cache (the Linux Page Cache) can be enabled or disabled.
These platform configurations are summarized in
Table~\ref{tab:hardware_platforms} (FC and SC stand for Fast Cache and Slow
cache, respectively, and FN and SN stand for Fast Network and Slow Network,
respectively).  We treat each of these configurations as a different
platform and perform simulation calibrations independently.  Because they
correspond to different ground truths with different hardware
configurations, they cannot be used as a larger aggregated dataset that can
be used for computing a single calibration. This is because our calibration
parameters pertain to the platform hardware characteristics, as described
hereafter.

\begin{table}[tbp]
  \centering
	\caption{Hardware Platform Configuration Specifications.}
	\label{tab:hardware_platforms}
	\begin{small}
	\begin{tabular}{|c|c|c|}
	\hline
	\textbf{Platform} & \textbf{RAM page cache} & \textbf{WAN interface} \\ \hline
	\HDDcFiber  &  disabled & 10 Gbps  \\ \hline
	\RAMcFiber & enabled & 10 Gbps    \\ \hline
	\HDDcDial  & disabled  & 1 Gbps   \\ \hline
	\RAMcDial & enabled & 1 Gbps    \\\hline
	\end{tabular}
	\end{small}
	\vspace*{-0.1in}
\end{table}

\medskip
\noindent{\bf Calibration Parameters --} In this case study the goal is to calibrate the
following four simulation parameters, which are depicted also in Figure~\ref{fig:platform}:
\begin{compactitem}
\item Compute node core speed (flop/sec);
\item Disk bandwidth (bit/sec);
\item LAN bandwidth (bit/sec); and
\item WAN bandwidth (bit/sec).
\end{compactitem}
Although the simulator takes as input core speed floating point operations
per second, these are better understood as work unit per second, where the
work unit is application-specific.  The simulator comes with two other
parameters (which we consider in Section\ref{sec:tradeoff}).  The SimGrid
and WRENCH frameworks provide simulation models configurable via hundreds
of parameters (for which we use the default values these frameworks
provide).  As a result, our parameter search space has low dimension, which
is why one can expect the simple algorithms in Section~\ref{sec:algs} to
perform well.

\medskip
\noindent{\bf Accuracy Metric --} 
The domain scientists intending to use this simulator to achieve the
objectives outlined in Section~\ref{sec:casestudy:context} have identified
performance metrics of interest. These are the the average job execution
times for each of the 3 compute nodes at the compute site and for each of
the 11 \hitrate values, for a total of 33 metrics. In this case study we
consider a single aggregate metric to quantify simulation accuracy: the
Mean Relative Error (MRE), in percentage, of these 33 metrics.  A lower MRE
value means a better accuracy.

\medskip
\noindent{\bf Domain Scientist Calibration --} In this case study we
quantify the improvement that automated calibration can bring w.r.t. a
calibration performed manually by a domain scientist.  This manual calibration
was performed by the second author of this paper. The approach was
incremental.  First, the core compute speeds were calibrated based on
ground-truth data obtained from \RAMcFiber (to 1,970 Mflops), so as to
minimize the overhead of network and I/O operation.  Second, the external
network bandwidth was calibrated for \HDDcDial and \RAMcDial (to
1.15~Gbps).  For \HDDcFiber and \RAMcFiber, it was assumed that the same
ratio of effective bandwidth to hardware-specified bandwidth applies, and
thus the external network bandwidth was set to 11.5~Gbps (10x higher
than for \HDDcDial and \RAMcDial).  Third, the
bandwidth of the HDD caches was calibrated based on \HDDcFiber (to 17
MBps).  Some other parameter values were not calibrated, but their values
were determined based on knowledge of the platform and simple benchmarks
(and variations of these values were observed to have negligible impact on
the simulation).  These parameter values are: the internal network
bandwidth (set to be 10~Gbps) and the Linux page cache speed (set to
1~GBps).

The manual calibration was done by starting from an initial guess
(based on hardware specifications) and searching for a value in a
neighborhood of this guess until a good match between the ground-truth data
and the simulated data was observed.  For the network bandwidths, the ground
truth data exhibits low variance across job execution times, and there is
negligible difference between the ground-truth data and the data obtained
in simulation when using the calibrated parameters. For the HDD cache
speed, there is higher variance across job execution times, especially at
high \hitrate, due to more concurrent HDD reads to the cache. HDD effects
(e.g., seek times) are not modeled by the simulator, and as a result the
simulator does not produce the same variance. The calibration was performed
to match the simulated data to the average of the ground-truth data. All
details on this manual calibration procedure, which we denote as \human,
are available in~\cite{max_thesis}.

\medskip
\noindent{\bf Parameter Ranges --} As explained in Section~\ref{sec:pb},
our automated calibration procedure requires that a range of possible
values be specified for each parameter.  In this case study, all parameters
are given the same $2^{20}$ - $2^{36}$ range.  This range was
determined loosely based on reasonable expectations regarding hardware
specifications, assuming that no specific knowledge of the platform is
available.


\begin{table}[h!]
  \centering
	\caption{MRE for calibration methods and platforms.}
	\label{tab:human}
	\begin{tabular}{|l|l|c|c|c|c|}
	\cline{3-6}
		\multicolumn{2}{l}{} & \multicolumn{4}{|c|}{\textbf{Platform}}    \\ \cline{3-6}
		\multicolumn{2}{c}{ }& \multicolumn{1}{|c|}{\HDDcFiber}  & \RAMcFiber & \HDDcDial  & \RAMcDial\\
		\cline{1-6}
		\multirow{4}{*}{\rotatebox[origin=c]{90}{{\bf Method~~~~}}}	
		& \human  & 23.21\%& 274.20\% & 18.48\% & 196.24\% \\ \cline{2-6}
		& \random & 22.07\%& 1.02\% & 14.69\% & 4.20\% \\ \cline{2-6}
		& \grid   & 24.10\%& 3.08\% & 16.72\% & 8.48\% \\ \cline{2-6}
		& \fixed  & 22.90\%& 1.50\% & 15.83\% & 6.59\% \\ \cline{1-6}
	\end{tabular}
\end{table}

\subsection{Results}
\label{sec:results}

\subsubsection{Simulation Accuracy}
\label{sec:accuracy}

\begin{table*}[btp]
	\centering
	\caption{Calibrated parameter values for platform \HDDcDial.}
	\label{tab:calibrated_values}
	\begin{tabular}{|l|c|c|c|c|}
		\cline{2-5}
		\multicolumn{1}{l|}{} & Core speed & Disk bandwidth & LAN bandwidth & WAN bandwidth\\
		\hline
		\human  & 1,970 Mflops & 17 MBps & 10.0 Gbps  & 1.15 Gbps \\ 
		\hline
		\random & 823 Mflops  & 17 MBps & 6.1 Gbps & 21.0 Gbps \\
		\hline
		\grid   & 1,073 Mflops & 17 MBps & 17.0 Gbps  & 0.27 Gbps \\
		\hline
		\fixed  & 778 Mflops  & 16 MBps & 2.5 Gbps & 57.0 Gbps \\
		\hline
		\end{tabular}
\end{table*}

Table~\ref{tab:human} shows MRE values for all calibration methods.   The
first observation is that the automated calibration methods almost always
improve on \human for all platforms.  The exception is \grid for the
\HDDcFiber platform, which leads to MRE higher than \human by less than a
point. The improvement is by only a few points for \HDDcFiber
and \HDDcDial, but more than 150 points for \RAMcFiber and \RAMcDial. For
these last two platform configurations recall that the manual calibration
procedure simply assumes the value for the Linux page cache speed to be
1~GBps (see Section~\ref{sec:methodology}.  This is likely the cause for
the high MRE values, as the automated calibration methods compute values
that are higher by $\sim$10x.  The poorest performing algorithm is,
expectedly, \grid.   But we note that all our algorithms lead to similar
MRE. This is because our search space is of low dimensionality and even
simple algorithms, such as \grid and \random, are able to find a good
calibration. \fixed is not significantly more effective because 
the objective function is
``mostly flat" along several dimensions (for parameters that do not
pertain to a bottleneck resource, as explained in the next section).

\subsubsection{Bottleneck Resources}
\label{sec:bottleneck}

Although our algorithms lead to similar simulation accuracy, they actually
compute quite different calibrations.  Table~\ref{tab:calibrated_values}
shows calibrated parameter values computed by all calibration methods for
platform \HDDcDial. We observe that all methods compute very
similar values for the disk bandwidth parameter (between 16 and 17~MBps).
However, for some of the other parameters, the computed values can be
wildly different. For instance, the values for the WAN bandwidth range from
0.27~Gbps to 57~Gbps, while the actual value is likely around 1Gbps.  The
reason is that the performance of the workload whose execution is simulated
is driven by a single bottleneck resource. Parameter values
pertaining to other resources thus have little impact on the simulated
execution. In \HDDcDial, the bottleneck for our ground-truth workload is
the disk because the Linux page cache is disabled.  The same observation
can be made for the other three platforms, where all algorithms compute
almost the same parameter value for the relevant bottleneck.  Note that the
\human calibrations (for each platform) have values that are likely more
accurate for non-bottleneck resource parameters due to the incremental
manual calibration approach described in Section~\ref{sec:methodology},
which benefits from specific knowledge that the domain scientist has
regarding platform configurations.

The problem with the above is that the computed calibrations are not
generalizable to all application workloads.  That is, the calibrated
simulator is valid only to simulate the execution of workloads that would
experience the same performance bottleneck as the ground-truth workload.
Specifically, our calibrated simulator (using any of our algorithms), is
only valid for simulating the execution of workloads with the same ratio of
compute to data volumes as the ground-truth workload. For these
workloads, the simulator is useful as it produces valid results for
simulating configurations with more or fewer jobs, with more of fewer
compute nodes and/or cores, and with different \hitrate values.

There are two solutions, which can be combined, for making calibrations
generalizable to workloads with a range of ratios of compute to data
volumes.  The first solution would be the use of more ground-truth
data obtained from the execution of workloads with different enough such
ratios that they experience different bottlenecks when executed on the same
platform.  In this case study, however, the domain scientist
collected ground-truth data for only one workload.  The reasons are:
(i)~producing a calibrated simulator for this workload only is still
useful, as explained above; and (ii)~collecting ground-truth data is
labor-, time, and energy-consuming. To allow for a fair comparison between
the \human calibration and our automated calibration approach, our
calibration algorithms use that same ground-truth data.  The second
solution would entail defining and using a simulation accuracy metric
whose value is not solely driven by parameter values that pertain to the
bottleneck resource.  The metric used in this case study (defined in
Section~\ref{sec:methodology}) is an aggregate metric that does not capture
the temporal structure of the workload's execution, but only takes into
account average job execution times. As a result, the durations of
activities (computations, I/O operations, network communications) that do
not execute on the bottleneck resources (but execute concurrently with
activities that do execute on these resources) have no impact on
simulation accuracy. A metric that captures the duration of these
activities would instead force the calibration algorithms to calibrate more
than just the parameters that pertain to bottleneck resources. Such a
metric could include, for instance, a measure of the discrepancy between
the start and/or end times of all data transfers, I/O operations, and
computations.

The questions of which level of diversity of ground-truth data is necessary
and which accuracy metric is sufficient to ensure that automated
calibration algorithms produce generalizable calibrations are beyond the
scope of this paper and left for future work. Answering these questions
will require conducting multiple case studies for different application
domains and collecting new ground-truth data. 

\subsubsection{Using Less Ground-Truth Data}

As seen in the previous section, more diverse ground-truth data is needed
to obtain calibrations that are valid for the full spectrum of
application workload configurations.  Given our available ground-truth data
in this case study, in all that follows we limit our scope to workloads
that have the same compute-data ratio as that of the ground-truth workload.
But within this scope, one may then wonder whether good calibrations can be
computed automatically using less ground-truth data.  Specifically, an
interesting question is whether good calibrations can be computed based on
only a subset of the \hitrate values.  If using less ground-truth data is
feasible, then the result is time, labor, and energy savings for the
overall simulation calibration procedure.

%
%

\begin{table}[htbp]
  \centering
	\begin{small}
	\caption{Best, median, and worst MRE when calibrating using subsets of the \hitrate values using \fixed for platform \RAMcDial.}
	\label{tab:overfitting}
	\begin{tabular}{|c|c|c|c|c|}
	\hline
	\textbf{\# \hitrate values }  & \textbf{\# Subsets} & \textbf{Best} & \textbf{Median} & \textbf{Worst}  \\ \hline
	1 & 5 &  20.51\% & 52.00\% & 7008.44\%  \\ \hline
	2 & 10 & 4.20\%  & 5.52\% & 21.10\%  \\ \hline
	3 & 10 & 4.20\%  & 4.20\% & 10.02\%  \\ \hline
	11 & 1 & 6.59\%  & 6.59\% & 6.59\%  \\ \hline
	\end{tabular}
	\end{small}
\end{table}

To answer the above question we use one of our algorithms (\fixed) to compute
calibrations when using subsets of a 5-element set of \hitrate values
$\{0.0, 0.3, 0.5, 0.7, 1.0\}$.  Table~\ref{tab:overfitting} shows results
for calibrating using all five 1-element subsets, all ten 2-element
subsets, and all ten 3-element subsets.  The last row of the table shows
results when using all 11 \hitrate values for calibration, as done in the
previous section.  These results are for platform \RAMcDial 
(results are similar for other platforms).  We do not show results
for each individual subset, but show instead the best, median, and worst
MRE values over all subsets with the same cardinality.

When calibrating using one \hitrate value the MRE range is
20.51\%-7,008.44\%, with the worst MREs (higher than 5,000\%) obtained when
calibrating based on one of the two extreme \hitrate values 0.0 and 1.0.
This is expected as with these values the caching behavior is markedly
different than that with intermediate values.  Calibrating using only one of these
intermediate values leads to MRE between 20.5\% and 52.0\%, with the lowest MRE
values achieved when calibrating based on \hitrate 0.5.  When using two
\hitrate values the MRE improves drastically. Although the worst MRE is at
21.10\%, the median is at 5.52\%.  In fact, only one of the ten possible
subsets leads to MRE above 9\%, for subset $\{0, 0.3\}$.  When
using three \hitrate values, the worst MRE improves to 10.02\%. Here again,
the median is equal to the best. Only subset $\{0, 0.3, 0.5\}$ leads to MRE
above 7\%. In these results, when using 2- or 3-element subsets, the
worst performing subset is always the one that includes only the smallest
\hitrate value. As long as there is reasonable diversity  in \hitrate
values, e.g., some below 0.5 and some above 0.5, the automated calibration
based on two or three \hitrate values leads to accuracy on par with that
obtained when calibrating with all 11 \hitrate values. 

Calibrating using $n$ \hitrate values can lead to better accuracy than
calibrating with $n' > n$ \hitrate values, i.e., using less ground-truth
data.  For instance, the best MRE value when using all 5 \hitrate values is
6.59\%, but is 4.20\% when using 2 or 3 \hitrate values.  The main reason
is that the same amount of time $T$ is allotted to the calibration
procedure regardless of how many \hitrate values are used.  Evaluating the
accuracy of a calibration requires $n'/n$ fewer simulator invocations when
using $n$ \hitrate values as opposed to $n'$ \hitrate values. Thus using
fewer \hitrate values makes it possible to explore the parameter space more
thoroughly within time $T$.  



We conclude that as long as a reasonably diverse set of ground-truth data
is used, good calibrations can be computed even when using relatively few values.
Nevertheless, in the following sections results are presented for
calibrations computed using all 11 \hitrate values.

\subsubsection{Trade-off between Speed and Accuracy}
\label{sec:tradeoff}

Our simulator takes as input more parameters than
the four
 that we have calibrated in previous sections.  We now consider
two specific additional parameters. The first parameter is the XRootD block
size, $B$. Each file in XRootD, like in most storage systems, is
partitioned into blocks. The jobs in the workload process input files block
by block, so that reading and processing data is done in a pipelined
fashion.  The second parameter is the buffer size, $b$, which specifies the
internal buffer size used by a storage service, for the purpose of
pipelining I/O and network operations, as done in production storage
systems.

The $B$ and $b$ parameters correspond to software
configuration parameters in the real-world system. Picking realistic
values for them could be done by inspecting the system's
configuration files. This was not done for the \platform platform
for this case study but, regardless, values are likely a few MBs or on the
order of KBs (e.g., the default XRootD block size is 2MB).
These parameters drive the number of discrete events
that must be simulated.  Given a job in the workload that needs to process
$s$ bytes of data, the number of simulated events for this 
job's execution is $O(s/B  + s/b))$. If $B$ and/or $b$ are low
relative to $s$, the simulation time can become prohibitively high.  
Hence, for these two parameters the goal is not to find
values that are as realistic as possible. Instead, the goal is to set
the values of these parameters so that the simulation time is below
some user-defined threshold and then calibrate all other parameters
automatically. The question is whether this calibration can still
lead to good simulation accuracy.  In other words, can the automated
calibration of the other parameters compensate for the potential loss
of accuracy due to simulating the execution at a higher granularity
(i.e., larger block and buffer sizes) than the real-world system?  

To answer this question we consider four combinations of $B$ and $b$
values, so that the average simulation time is $\sim$1 sec ($B=10^{10}$
byte, $b=10^8$ bytes), $\sim$3 sec ($B=10^9$ bytes, $b=10^7$ bytes),
$\sim$30 sec ($B=10^8$ bytes, $b=10^6$ bytes), or $\sim$5 min
($B=10^7$ bytes, $b=10^5$ bytes). We then run our automated calibration
procedure for platform \RAMcDial for each of our algorithms. In
all other sections we use $B=10^8$ bytes and $b=10^6$
bytes (for $\sim30$-sec simulation times).

\begin{table}[htbp]
  \centering
	\caption{MRE vs. average simulation time for platform \RAMcDial.}
	\label{tab:block_table}
	\begin{small}
	\begin{tabular}{|l|c|c|c|}
	\hline
	\textbf{Sim. time}    &  \fixed  & \grid    & \random \\ \hline
	\textbf{~$\sim$1 sec}  & 3.13\% & 4.50\%  & 2.93\% \\ \hline
	\textbf{~$\sim$3 sec}  & 4.26\% & 10.85\%  & 3.26\% \\ \hline
	\textbf{~$\sim$30 sec} & 6.59\% & 8.48\%  & 4.20\% \\ \hline
	\textbf{~$\sim$5 min}  & 13.58\% & 28.33\%  & 4.02\% \\ \hline
	\end{tabular}
	\end{small}
	\vspace*{-0.1in}
\end{table}

Table~\ref{tab:block_table} shows MRE vs. average simulation time for our
three algorithms. In general we observe that MRE increases as the simulation
time increases.  But there are some exceptions: \grid leads to higher
MRE with 3-sec simulation times than with 30-sec simulation times; and
\random leads to higher MRE with 30-sec simulation times than with 5-min
simulation times.  This may seem surprising given that all simulation parameter
values that are sampled when using the longer simulation time are also
sampled when using the shorter simulation time.  However, simulation
parameters that lead to low MRE for particular $B$ and $b$ values may lead
to high MRE for different $B$ and $b$ values.  For instance, the
best found calibration that achieves an MRE of 8.48\% with $B=10^8$ and
$b=10^6$ (30-sec simulation time), leads to an MRE of 30.42\% when used with
$B=10^9$ and $b=10^7$ (3-sec simulation time).

Regardless, the key observation is that for all algorithms the best MRE is
achieved for the fastest simulation time, i.e., for the largest $B$ and $b$
values With larger $B$ and $b$ values the simulation has a much higher
granularity than the real-world system, which would seem to imply worse
accuracy. However, with these large values the simulation time is short,
meaning that the calibration procedure can better explore the parameter
space, allowing it to find parameter values that lead to better accuracy in
spite of the higher granularity.  A lower granularity in the real-world
system means better utilization of the hardware resources due to
finer-grain pipelining of I/O, network, and compute activities.  Pipelining
thus increases the effective speed of I/O, network, and compute resources
for each job. In simulation this same increase can be achieved instead by
using a higher granularity and at the same time increasing the speed of the
corresponding simulated hardware resources, at least within some bounds. In
this case study, doing so allows the user to obtain a calibrated simulator
that is both fast and accurate. Because our calibration procedure is
automated, it would be straightforward for users of the simulator to
explore the accuracy-speed design space to achieve whatever user-specific
trade-off is the most desirable. Making this determination manually would
be prohibitively labor-intensive.

\subsubsection{Impact of the Time Bound $T$}
\label{sec:timebound}

\begin{figure}[t]
    \centering
    \includegraphics[width=0.99\columnwidth]{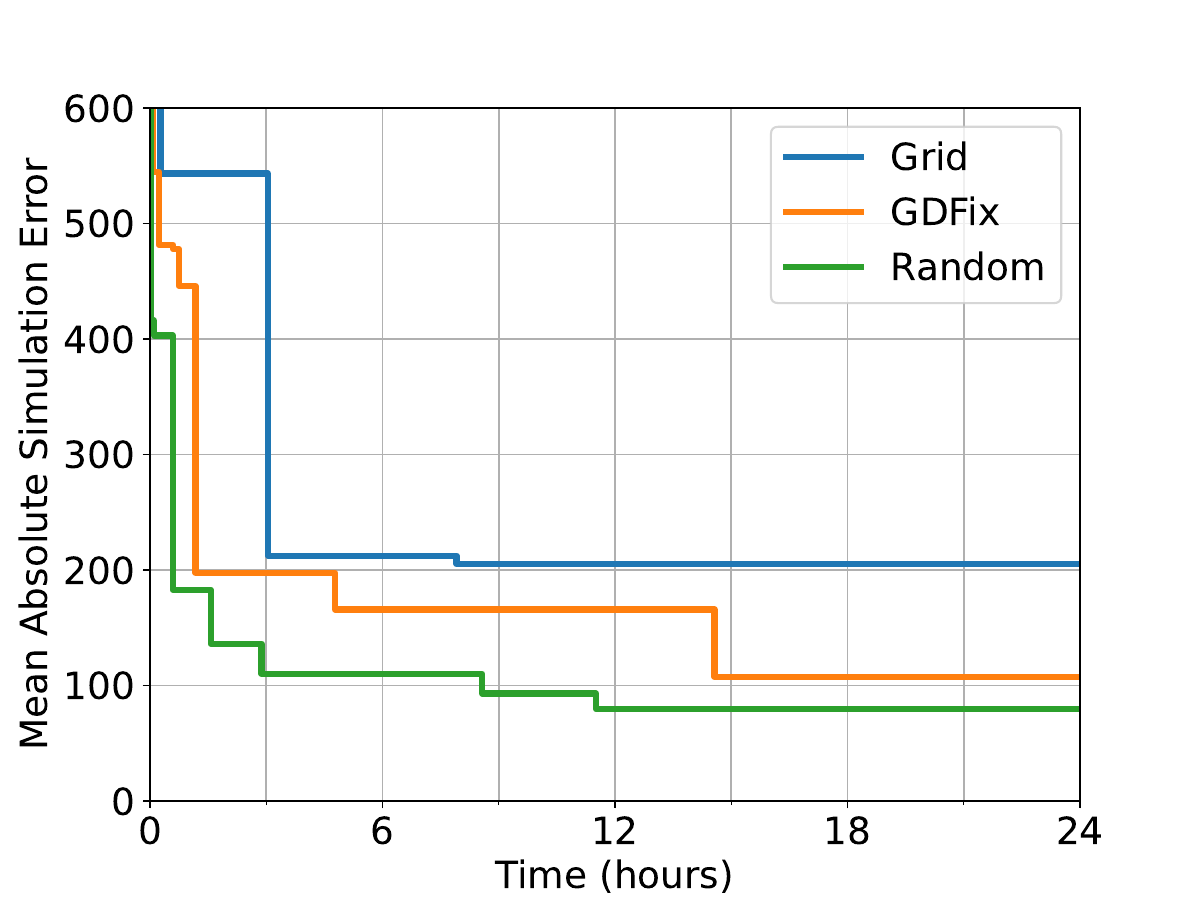}
	\caption{Absolute simulation error vs. time for platform \RAMcDial.}
    \label{fig:timebound}
\end{figure}

All previous results were obtained for an arbitrarily fixed
calibration time bound $T$ of 6 hours.  Figure~\ref{fig:timebound} plots
absolute simulation error vs. time, for time up to 24 hours, for platform
\RAMcDial (results are similar across all platforms). These results are
obtained using $B=10^8$ and $b=10^6$ bytes for the XRootD block size and
buffer size parameters, for which the simulation time is $\sim$30 sec.

As expected, all curves are non-increasing, with a very sharp initial
decrease. \grid leads to the worst results with the slowest convergence.
\random converges the most rapidly and leads to the lowest error overall.
\fixed is in between. Some improvements can be achieved by setting $T > 6$
hours, especially for \random and \fixed, with both algorithms converging
to similar error values.  Using a shorter $T \sim $3 hours would have
produced only marginally higher errors than with $T = 6$ hours.



%% file: sec-conclusion.tex
\section{Conclusion}
\label{sec:conclusion}

We have shown that the state of the art of PDC simulator calibration
comprises undocumented and/or labor-intensive ad-hoc approaches, which
motivates for developing automated calibration methods.  Via a case
study  from the field of High Energy Physics, we have demonstrated that
even simple algorithms can be used to improve upon a
calibration computed by a domain scientist. We have also shown that this
can be achieved with reduced amounts of ground-truth data, and that
automated calibration makes it possible to achieve desirable trade-offs
between simulation speed and simulation accuracy.

A clear future direction is to augment our case study and to perform case
studies for other PDC systems and simulators thereof. Doing so will allow
investigating which amount and diversity of ground-truth data, and which
accuracy metric definitions, are sufficient to compute calibrations that
are robust: the calibrated simulator should yield accurate results for the
full spectrum of possible application workload configurations.  In our case
study in this work we have kept the calibration parameter space at only 4
dimensions. But in practice, for this and other case studies, it could be
much larger with with hundreds of parameters. The simple algorithms we have
considered in this work will likely no longer be effective, and another
clear future direction is the use of Machine Learning algorithms.  In
particular, Bayesian Optimization is an attractive proposition as it is
highly effective for optimizing black-box functions that are relatively
expensive to evaluate, such as simulation accuracy metrics whose evaluation
entails invoking a simulator.
